\documentclass[12pt]{article}
\usepackage{mathrsfs}
\usepackage{amsmath,amssymb,latexsym,color,cancel,graphicx,bbm,colortbl}
\usepackage[english]{babel}
\usepackage[latin1]{inputenc}
\usepackage{ragged2e}
\begin{document}
\date{}

\title{$SU(1,1)$ coherent states for Dirac-Kepler-Coulomb problem in $D+1$ dimensions with scalar and vector potentials}
\author{D. Ojeda-Guill\'en$^{a}$,\footnote{{\it E-mail address:} dogphysics@gmail.com}\\ R. D. Mota$^{b}$ and V. D. Granados$^{a}$} \maketitle

\begin{minipage}{0.9\textwidth}
\small $^{a}$Escuela Superior de F{\'i}sica y Matem\'aticas,
Instituto Polit\'ecnico Nacional,
Ed. 9, Unidad Profesional Adolfo L\'opez Mateos, C.P. 07738, M\'exico D. F., Mexico.\\

\small $^{b}$Escuela Superior de Ingenier{\'i}a Mec\'anica y El\'ectrica, Unidad Culhuac\'an,
Instituto Polit\'ecnico Nacional, Av. Santa Ana No. 1000, Col. San
Francisco Culhuac\'an, Delegaci\'on Coyoac\'an, C.P. 04430, M\'exico D. F., Mexico.\\

\end{minipage}

\begin{abstract}
We decouple the Dirac's radial equations in $D+1$ dimensions with Coulomb-type scalar and vector potentials through appropriate transformations. We study each of these uncoupled second-order equations in an algebraic way by using an $su(1,1)$ algebra realization. Based on the theory of irreducible representations, we find the energy spectrum and the radial eigenfunctions. We construct the Perelomov coherent states for the Sturmian basis, which is the basis for the unitary irreducible representation of the $su(1,1)$ Lie algebra. The physical radial coherent states for our problem are obtained by applying the inverse original transformations to the Sturmian coherent states.

\end{abstract}


\section{Introduction}
Since Schr\"odinger introduced the harmonic oscillator coherent states \cite{Scrho}, they
have played a fundamental role in quantum mechanics. These coherent states are related to the
Heisenberg-Weyl group. The works of Barut \cite{BandG} and Perelomov \cite{Perel}
generalized the harmonic oscillator coherent states to those of any algebra of a symmetry group.

The coherent states have been obtained successfully for many problems,
reported in the references \cite{Klauderlibro}-\cite{Klimov}.
Related to the Perelomov coherent states for the $su(2)$ and $su(1,1)$ Lie
algebras, several works have been published, some of them are \cite{Eberly,BrifHO,BrifHO2}.

As one of the few exactly solvable problems in physics, the Kepler-Coulomb problem
has been treated in several ways, analytical \cite{THALL1}-\cite{ROBIN2}, factorization methods \cite{CHINOS1,CHINOS2},
shape-invariance \cite{LIMA}, SUSY QM for the first \cite{SUKU} and second-order \cite{JARVIS} differential equations,
two-variable realizations of the $su(2)$ Lie algebra \cite{MEX1} and using the Biedenharn-Temple operator \cite{HORVATHY}.
Its solubility is due to the conservation of the total angular momentum, and the Dirac
and Johnson-Lippmann operators \cite{DAHL}. In fact, is has been shown that the
supersymmetry charges are generated by the Johnson-Lippmann operator
\cite{DAHL}. Joseph was the first in studying the Kepler-Coulomb problem  in  $D+1$
dimensions by means of self-adjoint operators \cite{JOS}. The
energy spectrum and the eigenfunctions of this problem were obtained
by solving the confluent hypergeometric equation \cite{TUTIK,Dongpopov}.
Moreover, in \cite{JAPON} the Johnson-Lippmann operator for this
potential has been constructed and used to generate the SUSY
charges.

The Dirac equation with Coulomb-type vector and scalar potentials in $3+1$ dimensions
has been solved by using SUSY QM \cite{JUG} and the matrix form of SUSY QM based on intertwining operators \cite{LEVIATAN}. For the $D+1$-dimensional case it was treated by reducing the uncoupled radial second-order equations to those of the confluent hypergeometric
functions \cite{TUTIK,Dongpopov}. In  recent works, it has been studied the Dirac equation
for the three-dimensional Kepler-Coulomb problem \cite{JPA}, and with Coulomb-type
scalar and vector potentials in $D+1$ dimensions from an $su(1,1)$ algebraic approach \cite{Epl}.
Also, a Johnson-Lippmann operator has been constructed for  Coulomb-type scalar and vector potential in general spatial dimensions.
It was used to generate the SUSY charges \cite{DANIEL}.

In the relativistic regimen the spectrum of the Klein-Gordon Coulomb problem was calculated by using the $SO(2,1)$ coherent-state theory \cite{BOSCHI}. For the Dirac problem, only the coherent states for the three-dimensional relativistic Kepler-Coulomb potential have been treated \cite{DRAGA}. This study was based on the fact that the uncoupled second-order differential equations admits a two-variable $SU(2)$ symmetry. The additional variable to the radial coordinate was needed in order to close the $su(2)$ Lie algebra \cite{MEX1}. We notice that in references \cite{BOSCHI,DRAGA} the explicit closed form of the relativistic coherent states have not been obtained. However, there are no works on relativistic coherent states in which scalar and vector potentials are considered, in three or higher spatial dimensions.

The aim of the present work is to construct the radial $SU(1,1)$ Perelomov coherent states for the relativistic Kepler-Coulomb problem in $D+1$ dimensions with Coulomb-type scalar and vector potentials. Our treatment is restricted for bound states. We decouple the first-order Dirac equations to obtain the second-order differential equations. Each of these equations is written in terms of a set of radial operators which close the $su(1,1)$ Lie algebra. One of these operators is the so-called scaling operator. By an appropriate choice of the scaling parameter we diagonalize the second-order radial equation. We use the theory of unitary representations to find the energy spectrum and the radial wave functions from the Sturmian basis (group basis). We construct the $SU(1,1)$ Perelomov coherent states for the Sturmian basis and the inverse transformations are applied to these states to obtain the radial $SU(1,1)$ Perelomov coherent states for the relativistic Kepler-Coulomb problem.

This work is organized as follows. In section $2$ we obtain the uncoupled second-order differential equations satisfied by the radial components. The $su(1,1)$ Lie algebra generators for the uncoupled second-order differential equations are introduced. The energy spectrum and radial wave functions are found. In section $3$, we obtain the explicit expression of $SU(1,1)$ Perelomov coherent states for the relativistic Kepler-Coulomb problem in $D+1$ dimensions with scalar and vector potentials. Finally, we give some concluding remarks.

\section{Second order radial equations}

The Dirac equation in $D+1$ dimensions for a central field is given by \cite{Dongpopov}
\begin{equation}\label{Dir}
i\frac{\partial\Psi}{\partial{t}}=H\Psi,\qquad
H=\sum^D_{a=1}\alpha_a{p}_a+\beta\left(m+V_s\left(r\right)\right)+V_v\left(r\right),
\end{equation}
with  $\hbar=c=1$, $m$ is the mass of the particle, $V_s$ and $V_v$ are the
spherically symmetric scalar and vector potentials, respectively and
\begin{equation}
p_a=-i\partial_a=-i\frac{\partial}{\partial{x}_a}\hspace{0.5cm}
1\leq a\leq D.
\end{equation} In (\ref{Dir}), $\alpha_a$ and $\beta$ satisfy the anticommutation relations
\begin{align}\nonumber
\alpha_a\alpha_b+\alpha_b\alpha_a=2\delta_{ab}\textbf{1},\\
\alpha_a\beta+\beta\alpha_a=0,\\\nonumber
\alpha_a^2=\beta^2=\texttt{1}.
\end{align}

In $D$ spatial dimensions, the orbital angular momentum operators
$L_{ab}$, the spinor operators $S_{ab}$ and the total angular
momentum operators $J_{ab}$ are defined as
\begin{eqnarray}\nonumber
L_{ab}=-L_{ba}=ix_a\frac{\partial}{\partial
x_b}-ix_b\frac{\partial}{\partial x_a},\hspace{0.3cm}
S_{ab}=-S_{ba}=i\frac{\alpha_a\alpha_b}{2},\hspace{0.3cm}
J_{ab}=L_{ab}+S_{ab}.\\
L^2=\sum_{a<b}^D L^2_{ab},\hspace{0.3cm} S^2=\sum_{a<b}^D
S^2_{ab},\hspace{0.3cm} J^2=\sum_{a<b}^D J^2_{ab},\hspace{0.3cm}
1\leq a\leq b\leq D.
\end{eqnarray}
Hence, for a spherically symmetric potential, the total angular
momentum operator $J_{ab}$ and the spin-orbit operator
$K_D=-\beta\left(J^2-L^2-S^2+\frac{\left(D-1\right)}{2}\right)$
commute with the Dirac Hamiltonian. For a given total angular
momentum $j$, the eigenvalues of the operator $K_D$ are
$\kappa_D=\pm\left(j+\left(D-2\right)/2\right)$, where the minus
sign is for aligned spin $j=\ell+\frac{1}{2}$, and the plus sign is
for unaligned spin $j=\ell-\frac{1}{2}$.

We propose the Dirac wave function of equation (\ref{Dir}) to be of
the form
\begin{equation}\label{SolDir}
\Psi(\vec r, t)=r^{-\frac{D-1}{2}}
\begin{pmatrix}
F_{\kappa_D}\left(r\right)Y_{jm}^\ell(\Omega_D)\\
iG_{\kappa_D}\left(r\right)Y_{jm}^{\ell'}(\Omega_D)
\end{pmatrix}e^{-iEt},
\end{equation}
being $F_{k_D}(r)$ and $G_{k_D}(r)$ the radial functions,
$Y_{jm}^\ell(\Omega_D)$ and $Y_{jm}^{\ell'}(\Omega_D)$ the
hyperspherical harmonic functions coupled with the total angular
momentum quantum number $j$, and $E$ the energy. Thus, equation
(\ref{Dir}) leads to the radial equations
\begin{equation}\label{difGG}
\begin{pmatrix}
\frac{dF{\kappa_D}}{dr}\\
\frac{dG_{\kappa_D}}{dr}
\end{pmatrix}=
\begin{pmatrix}
-\frac{\kappa_D}{r} & V_s-V_v+m+E \\
V_v+V_s+m-E & \frac{\kappa_D}{r} \end{pmatrix}
\begin{pmatrix}
F_{\kappa_D}\\
G_{\kappa_D}
\end{pmatrix}.
\end{equation}

We consider the Coulomb-type scalar and vector potentials
\begin{equation}\label{Poten}
V_v= -\frac{\alpha_v}{r},\hspace{2cm} V_s=-\frac{\alpha_s}{r},
\end{equation}
with $\alpha_v$ and $\alpha_s$ positive constants. Therefore, from
equation (\ref{difGG}) we obtain
\begin{equation}\label{dg}
\begin{pmatrix}
\frac{dF_{k_D}}{dr}\\
\frac{dG_{k_D}}{dr}
\end{pmatrix}+
\frac{1}{r}\begin{pmatrix}
k_D & \alpha_v-\alpha_s \\
\alpha_v+\alpha_s & -k_D \end{pmatrix}
\begin{pmatrix}
F_{k_D}\\
G_{k_D}
\end{pmatrix}=\begin{pmatrix}
0 & m+E\\
m-E & 0 \end{pmatrix}\begin{pmatrix}
F_{k_D}\\
G_{k_D}
\end{pmatrix}.
\end{equation}

Now, we find a matrix $M$ which satisfies the property
\begin{equation}
M^{-1}\begin{pmatrix}
k_D & \alpha_v-\alpha_s \\
\alpha_v+\alpha_s & -k_D \end{pmatrix}M=
\begin{pmatrix}
-s & 0 \\
0 & s\end{pmatrix},
\end{equation}
where $s=\sqrt{k_D^2-\alpha_+\alpha_-}$, $\alpha_+=\alpha_v+\alpha_s$ and  $\alpha_-=\alpha_v-\alpha_s$.
This requirement leads to obtain
 \begin{equation}
M=\begin{pmatrix}
s-k_D & -(\alpha_v-\alpha_s) \\
-(\alpha_v+\alpha_s)& s-k_D \end{pmatrix}.
\end{equation}
This matrix and the definition $\begin{pmatrix}
u_{k_D}\\
v_{k_D}
\end{pmatrix}=M^{-1}\begin{pmatrix}
F_{k_D}\\
G_{k_D}
\end{pmatrix}$ allow to write the Dirac equation (\ref{dg})
as
\begin{equation}
\begin{pmatrix}
-(E-m)-\frac{\alpha_+(\alpha_v E+\alpha_s m)}{s(s-k)} &- \frac{d}{dr}-\frac{s}{r}+\frac{\alpha_v E+\alpha_s m}{s} \\
\frac{d}{dr}-\frac{s}{r}+\frac{\alpha_v E+\alpha_s m}{s} & -(E+m)-\frac{\alpha_-(\alpha_v E+\alpha_s m)}{s(s-k)} \end{pmatrix}
\begin{pmatrix}
u_{k_D}\\
v_{k_D}
\end{pmatrix}=0.\label{acopladas}
\end{equation}

We decouple this matrix equation to obtain the Dirac's second-order equations
\begin{equation}
\begin{pmatrix}
-\frac{d^2}{d r^2}+\frac{s(s - 1)}{r^2}-\frac{2(\alpha_v E+\alpha_s M)}{r}+M^2-E^2 & 0 \\
0 & -\frac{d^2}{d r^2}+\frac{s(s+1)}{r^2}-\frac{2(\alpha_v E+\alpha_s M)}{r}+M^2-E^2
\end{pmatrix}
\begin{pmatrix}
u_{k_D}\\
v_{k_D}
\end{pmatrix}=0.\label{desacopladas}
\end{equation}
Notice that these equations are of the same mathematical form. We can get the second one from the first one by substituting $s\rightarrow s+1$.
Also, these equations reduce to those for the Kepler-Coulomb problem in the three-dimensional space with vector potential only \cite{THALL2} and to those with scalar and vector potentials \cite{LEVIATAN}.

\subsection{$SU(1,1)$ radial coherent states}

We consider the operators
\begin{align}
A_0&=\frac{1}{2}\left(rP_r^2+\frac{s(s+1)}{r}+r\right),\label{HA1}\\
A_1&=\frac{1}{2}\left(rP_r^2+\frac{s(s+1)}{r}-r\right),\label{HA2}\\
A_2&=rP_r=-i r\left(\frac{d}{d r}+\frac{1}{r}\right)\label{HA3},
\end{align}
where the operator $P_r^2$ is defined by
\begin{equation}
P_r^2=-\frac{d^2}{d r^2}-\frac{2}{r}\frac{d}{d r}.
\end{equation}
These operators close the $su(1,1)$ Lie algebra \cite{BARUT,HECHT} of equation (\ref{com}) of Appendix. Due to the similarity of the uncoupled equations (\ref{desacopladas}), hereafter we shall focus just on the radial Dirac equation
for the eigenfunction $v_{k_D}$. It can be written as
\begin{equation}
\left(-\frac{d^2}{d r^2}-\frac{2}{r}\frac{d}{d r}+\frac{s(s+1)}{r^2}-\frac{2(\alpha_v E+\alpha_s M)}{r}+M^2-E^2\right){\tilde v(r)}=0,\label{Abajo}
\end{equation}
where we have defined $v(r)=r{\tilde v(r)}$, or equivalently as
\begin{equation}
{\tilde \Omega}(E) |{\tilde v (r)}\rangle=0,\label{crusial}
\end{equation}
where $|{\tilde v (r)}\rangle$ represents the physical states and
\begin{equation}
{\tilde \Omega}(E) \equiv \left(-\frac{d^2}{d r^2}-\frac{2}{r}\frac{d}{d r}+\frac{s(s+1)}{r^2}-\frac{2(\alpha_v E+\alpha_s M)}{r}+M^2-E^2\right). \label{ham}
\end{equation}
By multiplying equation (\ref{crusial}) by $\frac{1}{2}r$, it can be rewritten in terms of
the operators (\ref{HA1})-(\ref{HA3}) as
\begin{equation}
\left(\frac{1}{2}(A_0+A_1)+\frac{1}{2}(M^2-E^2)(A_0-A_1)-(\alpha_v E+\alpha_s M)\right){\tilde v(r)}=0.\label{EIGEN}
\end{equation}

If we introduce the scaling operator $e^{i\theta A_2}$, with the Baker-Campbell-Hausdorff formula we can show that
\begin{align}
e^{-i\theta A_2}A_0 e^{+i\theta A_2}=A_0 \cosh(\theta)+A_1 \sinh(\theta),\label{clave1}\\
e^{-i\theta A_2}A_1e^{+i\theta A_2}=A_0 \sinh(\theta)+A_1 \cosh(\theta),\label{clave2}
\end{align}
From these equations it follows that
\begin{equation}
e^{-i\theta A_2}(A_0 \pm A_1)e^{i\theta A_2}=e^{\pm \theta}(A_0\pm A_1).\label{paso}
\end{equation}
Therefore, by substituting (\ref{paso}) into equation (\ref{EIGEN}) we obtain
\begin{equation}
\Omega(E)|{\bar v}(r)\rangle=\left(\frac{1}{2}e^{\theta}(A_0+A_1)+\frac{1}{2}(M^2-E^2)e^{-\theta}(A_0-A_1)-(\alpha_v E+\alpha_s M)\right)|{\bar v}(r)\rangle=0,
\end{equation}
where  $|\bar{v}(r)\rangle=e^{-i\theta A_2}|\tilde{v}(r)\rangle $ and $\Omega(E)=e^{-i\theta A_2}{\tilde\Omega}(E) e^{i\theta A_2}$. By grouping common factors of $A_0$ and $A_1$, we
set $\theta$  such that the coefficient of $A_1$ vanishes. This leads to find
\begin{equation}
\theta=\ln(\sqrt{M^2-E^2}). \label{teta}
\end{equation}
Thus
\begin{equation}
\Omega(E)|{\bar v(r)}\rangle=\left(\sqrt{M^2-E^2} A_0-(\alpha_v E+\alpha_s M)\right)|{\bar v(r)}\rangle=0\label{}.
\end{equation}
If we take $|{\bar v(r)}\rangle$ as an $SU(1,1)$ group state $|{\bar n},{\bar s} \rangle$, this equation and equation (\ref{k0n}) of Appendix
allow us to obtain the energy spectrum
\begin{equation}
\frac{E_n}{m}=\frac{-\alpha_v \alpha_s+(n+s)\sqrt{(n+s)^2+\alpha_v^2-\alpha_s^2}}{\alpha_v^2+(n+s)^2}.\label{espectro}
\end{equation}
This result is in full agreement with the energy spectrum in general dimensions reported in references \cite{TUTIK,Dongpopov,Epl,DANIEL}. In the first two works the energy spectrum was obtained in an analytical way by means of confluent hypergeometric functions, while in the last two works the spectrum was obtained algebraically. In three dimensions, this result also reduce to the reported in \cite{LEVIATAN}.

The eigenfunctions basis for the irreducible unitary representations of the $su(1,1)$ Lie algebra with realization given by equations  (\ref{HA1})-(\ref{HA3}) (Sturmian basis) is \cite{HECHT,GERRY}
\begin{equation}
\langle r |{\bar n},{\bar s}\rangle={\bar v}_{n\;s}(r)=2\left[\frac{(n-1)!}{(n+2s)!}\right]^{1/2}(2r)^{s}e^{-r}L_{n-1}^{2s+1}(2r)\label{sturmv}.
\end{equation}
The uncoupled equations for $v_{k_D}$ and $u_{k_D}$ are related by making $s\rightarrow s-1$. However, since both eigenfunctions belong to the same energy level, equation (\ref{espectro}) imposes the change $n\rightarrow n+1$. Thus, the group state $|{\bar u(r)}\rangle$ is
\begin{equation}
{\bar u}_{n\;s}(r)=2\left[\frac{n!}{(n+2s-1)!}\right]^{1/2}(2r)^{s-1}e^{-r}L_{n}^{2s-1}(2r)\label{sturmu}.
\end{equation}
Therefore, the radial functions $F_{k_D}(r)$ and $G_{k_D}(r)$ can be obtained by means of the group states ${\bar u}(r)$ and ${\bar v}(r)$ as follows
\begin{equation}
\begin{pmatrix}
F_{k_D}\\
G_{k_D}
\end{pmatrix}=M\begin{pmatrix}
A_nu_{k_D}\\
B_nv_{k_D}
\end{pmatrix}=M\begin{pmatrix}
A_nr{\tilde u}\\
B_nr{\tilde v}
\end{pmatrix}=M\begin{pmatrix}
A_nre^{i\theta A_2}{\bar u}\\
B_nre^{i\theta A_2}{\bar v}
\end{pmatrix},\label{radiales}
\end{equation}
where $A_n$ and $B_n$ are two normalization constants to be determined. The physical states $|{\tilde u(r)}\rangle$ and $|{\tilde v(r)}\rangle$ are obtained by using \cite{HECHT}
\begin{equation}
e^{i\theta A_2}f(r)=e^{\theta}f(e^{\theta}r),
\end{equation}
being $f(r)$ an arbitrary spherically symmetric function. Thus, the physical radial eigenfunctions are
\begin{equation}
\tilde{u}(r)=A_n (2ar)^{s-1}e^{-ar}L_{n}^{2s-1}(2ar),\label{scalingu}
\end{equation}
\begin{equation}
\tilde{v}(r)=B_n (2ar)^{s}e^{-ar}L_{n-1}^{2s+1}(2ar),\label{scalingv}
\end{equation}
where $a=\sqrt{m^2-E^2}$. The coefficients $A_n$ and $B_n$ contain the normalization constants of the Sturmian basis functions and the relationship between these two coefficients can be obtained by evaluating the coupled equations (\ref{acopladas}) in the limit $r\rightarrow0$. If we define
\begin{equation}
\omega=-(E-m)-\frac{\alpha_+(\alpha_vE+\alpha_sm)}{s(s-k)},
\end{equation}
and use the formula
\begin{equation}
L_n^{\alpha}(0)=\frac{\Gamma(n+\alpha+1)}{n!\Gamma(\alpha+1)},
\end{equation}
the first of equations (\ref{acopladas}) leads to
\begin{equation}
B_n 2a(s+1)L_{n-1}^{2s+1}\left(0\right)=-B_n2asL_{n-1}^{2s+1}\left(0\right)+\omega A_nL_n^{2s-1}\left(0\right).
\end{equation}
Thus, the relationship between $A_n$ and $B_n$ is given by
\begin{equation}
B_n=\frac{\omega sA_n}{an(n+2s)}.
\end{equation}
Therefore, from equation (\ref{radiales}) the radial eigenfunctions $F_{k_D}$ and $G_{k_D}$ are given explicitly by
\begin{equation}
\begin{pmatrix}
F_{k_D}\\
G_{k_D}
\end{pmatrix}=A_n(2a)^{s-1}r^se^{-ar}
\begin{pmatrix}
F_1 L_{n}^{2s-1}(2ar)+F_2rL_{n-1}^{2s+1}(2ar)\\
G_1 L_{n}^{2s-1}(2ar)+G_2rL_{n-1}^{2s+1}(2ar)
\end{pmatrix},\label{radiales2}
\end{equation}
where $F_1=s-k$, $F_2=-2\omega s\frac{(\alpha_v-\alpha_s)}{n(n+2s)}$, $G_1=-(\alpha_v+\alpha_s)$ and $G_2=\frac{2\omega s(s-k)}{n(n+2s)}$.
Thus, this expression is the radial spinor for the Kepler-Coulomb problem in $D+1$ dimensions with scalar $-\alpha_s/r$ and vector potential $-\alpha_v/r$.

The coefficient $A_n$ can be explicitly calculated by using the relativistic normalization in $D$ spatial dimensions \cite{TUTIK}
\begin{equation}
\int_{0}^{\infty}\left(F^*F+G^*G\right)dr=1.\label{normalizacion}
\end{equation}
Since $F$ and $G$ are real functions, $F^*=F$ and $G^*=G$. These integrals can be computed by using the Laguerre integrals \cite{UVAROV}
\begin{equation}
\int_{0}^{\infty}e^{-2ar}r^{\alpha+1}\left[L_{n}^{\alpha}(2ar)\right]^2dr=\frac{(2n+\alpha+1)\Gamma(n\alpha+1)}{n!(2a)^{\alpha+2}},
\end{equation}
\begin{equation}
\int_{0}^{\infty}e^{-2ar}r^{\alpha}L_{n-1}^{\alpha}(2ar)L_{n}^{\alpha-2}(2ar)dr=\frac{-2\Gamma(n+\alpha)}{(n-1)!(2a)^{\alpha+1}}.
\end{equation}
With these results we obtain that the coefficient $A_n$ is given by
\begin{equation}
A_n=2\left(\frac{a^3n!}{\Gamma(n+2s)(n+s)(\sigma+\tau+\chi)}\right)^{1/2},
\end{equation}
where
\begin{align}
\sigma\equiv(s-k)^2+(\alpha_v+\alpha_s)^2,\quad\quad\tau\equiv\frac{4\omega s(\alpha_v-\alpha_s)(s-k)}{n+s},\nonumber\\
\chi\equiv\frac{\omega^2s^2((s-k)^2+(\alpha_v-\alpha_s)^2)}{n(n+2s)}.
\end{align}
In three-spatial dimensions and $\alpha_s=0$, above results are in full accordance with those reported in references \cite{THALL2} and \cite{UVAROV}.

\section{$SU(1,1)$ radial coherent states for the relativistic Kepler-Coulomb problem in $D+1$ dimensions}

Since each of the sets of functions $\bar{u}$ and $\bar{v}$ are a Sturmian basis for the $su(1,1)$ Lie algebra \cite{HECHT,GERRY}, we are able to generate the coherent states for these basis. We shall use the Sturmian basis coherent states and the transformation given by equation (\ref{radiales}) to construct the relativistic coherent states of the radial functions $F_{k_D}(r)$ and $G_{k_D}(r)$. The $SU(1,1)$ Perelomov coherent states are defined as \cite{Perellibro}
\begin{equation}
|\zeta\rangle=D(\xi)|k,0\rangle=(1-|\xi|^2)^k\sum_{s=0}^\infty\sqrt{\frac{\Gamma(n+2k)}{s!\Gamma(2k)}}\xi^s|k,s\rangle,
\end{equation}
 where $D(\xi)$ is the displacement operator and $|k,0\rangle$ is the lowest normalized state. We apply the operator $D(\xi)$ to the ground state of the functions $\bar{u}$ and $\bar{v}$ to obtain their $SU(1,1)$ coherent states
\begin{equation}
\bar{u}(r,\xi)=\frac{2(1-|\xi|^2)^s}{\sqrt{\Gamma(2s)}}e^{-r}(2r)^{s-1}\sum_{n=0}^\infty \xi^nL_n^{2s-1}(2r),
\end{equation}
\begin{equation}
\bar{v}(r,\xi)=\frac{2(1-|\xi|^2)^{s+1}}{\sqrt{\Gamma(2s+2)}}e^{-r}(2r)^{s}\sum_{n=1}^\infty \xi^{n-1}L_{n-1}^{2s+1}(2r).
\end{equation}
The sum of these coherent states can be computed by using the Laguerre polynomials generating function
\begin{equation}
\sum_{n=0}^\infty L_n^\nu(x)y^n=\frac{e^{xy/(1-y)}}{(1-y)^{\nu+1}}.
\end{equation}
By applying the scaling operator $e^{i\theta A_2}$, with the scaling parameter $\theta$ given by equation (\ref{teta}), and multiplying by $r$, we obtain the physical coherent states
\begin{equation}
u_{k_D}(r,\xi)=A_n'\frac{(1-|\xi|^2)^s}{\sqrt{\Gamma(2s)}}e^{-ar}a^{s-1}(2r)^s\frac{e^{-2ar\xi/(1-\xi)}}{(1-\xi)^{2s}},
\end{equation}
\begin{equation}
v_{k_D}(r,\xi)=B_n'\frac{(1-|\xi|^2)^{s+1}}{\sqrt{\Gamma(2s+2)}}e^{-ar}a^{s}(2r)^{s+1}\frac{e^{-2ar\xi/(1-\xi)}}{(1-\xi)^{2s+2}},
\end{equation}
where $A_n'$ and $B_n'$ are two normalization constants. Since $u_{k_D}(r,\xi)$ and $v_{k_D}(r,\xi)$ also satisfy the coupled equations (\ref{acopladas}), the relationship between these two relativistic normalization constants is obtained in a similar way as we proceeded at the end of the last section. Thus, in the limit $r\rightarrow0$ we obtain from the first coupled equation
\begin{equation}
B_n'=\frac{\omega(1-\xi)^2}{a(1-|\xi|^2)}\sqrt{\frac{s}{2(2s+1)}}A_n'.
\end{equation}
Thus, the $SU(1,1)$ radial coherent states $F_{k_D}$ and $G_{k_D}$ for the relativistic Kepler-Coulomb problem in $D+1$ dimensions with scalar and vectorial potentials are
\begin{equation}
\begin{pmatrix}
F_{k_D}(r,\xi)\\
G_{k_D}(r,\xi)
\end{pmatrix}=A_n'\frac{(1-|\xi|^2)^s(2r)^{s}a^{s-1}}{\sqrt{\Gamma(2s)}(1-\xi)^{2s}}e^{\frac{-ar(1+\xi)}{1-\xi}}
\begin{pmatrix}
s-k-\frac{(\alpha_v-\alpha_s)\omega r}{2s+1}\\
-(\alpha_v+\alpha_s)+\frac{(s-k)\omega r}{2s+1}
\end{pmatrix}.
\end{equation}
To obtain the normalization coefficient $A_n'$ we use the relativistic normalization, equation (\ref{normalizacion}). The integrals are reduced to an appropriate gamma function. This leads to obtain
\begin{equation}
A_n'=\left(\frac{a^3(1-|\xi|^2)}{s(1-\xi)(1-\xi^*)(\sigma'+\tau'+\chi')}\right)^{1/2},
\end{equation}
where
\begin{eqnarray}
\sigma'=(s-k)^2+(\alpha_v+\alpha_s)^2,\quad\tau'=-\frac{2\omega(s-k)\alpha_v(1-\xi)(1-\xi^*)\Gamma(2s+1)}{a(1-|\xi|^2)},\\
\chi'=\left(\frac{\omega(1-\xi)(1-\xi^*)}{(2s+1)a(1-|\xi|^2)}\right)^2\Gamma(2s+3)\left((\alpha_v-\alpha_s)^2+(s-k)^2\right).
\end{eqnarray}
Therefore, we constructed the $SU(1,1)$ coherent states for the Dirac-Kepler-Coulomb problem, in $D+1$ dimensions with scalar and vector potentials. As was pointed out in the Introduction, to our knowledge these states have not been calculated previously. Due to the generality of our calculation, we can hold the following particularly cases: a) In general dimensions with only one of the potentials or b) In three spatial dimensions with both or only one of the potentials.

\section{Appendix: $SU(1,1)$ Perelomov coherent states}

The $su(1,1)$ Lie algebra is spanned by the generators $K_{+}$, $K_{-}$
and $K_{0}$, which satisfy the commutation relations \cite{Vourdas}
\begin{eqnarray}
[K_{0},K_{\pm}]=\pm K_{\pm},\quad\quad [K_{-},K_{+}]=2K_{0}.\label{com}
\end{eqnarray}
The action of these operators on the Fock space states
$\{|k,n\rangle, n=0,1,2,...\}$ is
\begin{equation}
K_{+}|k,n\rangle=\sqrt{(n+1)(2k+n)}|k,n+1\rangle,\label{k+n}
\end{equation}
\begin{equation}
K_{-}|k,n\rangle=\sqrt{n(2k+n-1)}|k,n-1\rangle,\label{k-n}
\end{equation}
\begin{equation}
K_{0}|k,n\rangle=(k+n)|k,n\rangle,\label{k0n}
\end{equation}
where $|k,0\rangle$ is the lowest normalized state. The Casimir
operator $K^{2}=-K_{+}K_{-}+K_{0}(K_{0}-1)$ for any irreducible
representation satisfies $K^{2}=k(k-1)$. Thus, a representation of
$su(1,1)$ algebra is determined by the number $k$. For the purpose of the
present work we will restrict to the discrete series only, for which
$k>0$.

The $SU(1,1)$ Perelomov coherent states $|\zeta\rangle$ are
defined as \cite{Perellibro}
\begin{equation}
|\zeta\rangle=D(\xi)|k,0\rangle,\label{defPCS}
\end{equation}
where $D(\xi)=\exp(\xi K_{+}-\xi^{*}K_{-})$ is the displacement
operator and $\xi$ is a complex number. From the properties
$K^{\dag}_{+}=K_{-}$ and $K^{\dag}_{-}=K_{+}$ it can be shown that
the displacement operator possesses the property
\begin{equation}
D^{\dag}(\xi)=\exp(\xi^{*} K_{-}-\xi K_{+})=D(-\xi),
\end{equation}
and the so called normal form of the displacement operator is given by
\begin{equation}
D(\xi)=\exp(\zeta K_{+})\exp(\eta K_{0})\exp(-\zeta^*
K_{-})\label{normal},
\end{equation}
where $\xi=-\frac{1}{2}\tau e^{-i\varphi}$, $\zeta=-\tanh
(\frac{1}{2}\tau)e^{-i\varphi}$ and $\eta=-2\ln \cosh
|\xi|=\ln(1-|\zeta|^2)$ \cite{Gerry}. By using this normal form of the displacement
operator and equations (\ref{k+n})-(\ref{k0n}), the Perelomov coherent states are found to
be \cite{Perellibro}
\begin{equation}
|\zeta\rangle=(1-|\xi|^2)^k\sum_{s=0}^\infty\sqrt{\frac{\Gamma(n+2k)}{s!\Gamma(2k)}}\xi^s|k,s\rangle.\label{PCN}
\end{equation}

\section{Concluding remarks}

We constructed the $SU(1,1)$ Perelomov coherent states for the most general case of the Dirac's equation with Coulomb-type scalar and vector potentials in $D+1$ dimensions. We applied a series of transformations to the original Hamiltonian to obtain the uncoupled second-order differential equations satisfied by the radial components. We wrote each of the radial equations as a linear combination of the $su(1,1)$ Lie algebra generators. The irreducible representation theory allowed to obtain the energy spectrum and the radial eigenfunctions in a purely algebraic way.

We obtained the Perelomov coherent states for the irreducible representation basis of the $su(1,1)$ algebra (Sturmian basis). The physical radial coherent states for our problem were constructed by applying the inverse original transformations to the Sturmian coherent states. Notice that in our treatment we did not introduce the rescaling radial variable (which includes the energy), commonly used in this problem \cite{Dongpopov,Epl}. This allowed to use the $su(1,1)$ irreducible representation theory by means of the Sturmian basis to construct the relativistic coherent states. 

\section*{Acknowledgments}
This work was partially supported by SNI-M\'exico, COFAA-IPN,
EDI-IPN, EDD-IPN, SIP-IPN project number $20130620$.

\end{document}